\documentclass[12pt]{report}

\usepackage{graphicx}
\usepackage{amssymb}
\usepackage{multirow}
\usepackage{multicol}
\usepackage[square,numbers]{natbib}
\usepackage{longtable}
\usepackage{fancyhdr}
\usepackage{tikz}
\usepackage[tikz]{mdframed}
\usepackage{url}

\usepackage{color, soul}

\usepackage{titlesec} 
\usepackage[left=4cm,right=4cm,top=2.5cm,bottom=4cm]{geometry}

\definecolor{DarkRed}{RGB}{163,22,31}
\definecolor{LightRed}{RGB}{247,14,45}
\definecolor{PaleRed}{RGB}{239,134,148}
\definecolor{cpiGray}{RGB}{106,100,100}

\titleformat*{\section}{\color{DarkRed}\normalfont\bfseries\LARGE}
\titleformat*{\subsection}{\color{LightRed}\normalfont\bfseries\LARGE}
\titleformat*{\subsubsection}{\color{PaleRed}\normalfont\bfseries\large}

\title{A Survey of Relevant Text Mining Technology}
\author{Claudia Peersman, Matthew Edwards, Emma Williams, \& Awais Rashid}

\renewcommand{\maketitle}{\newpage
\newgeometry{margin = 0in}
\includegraphics[width=3.09in]{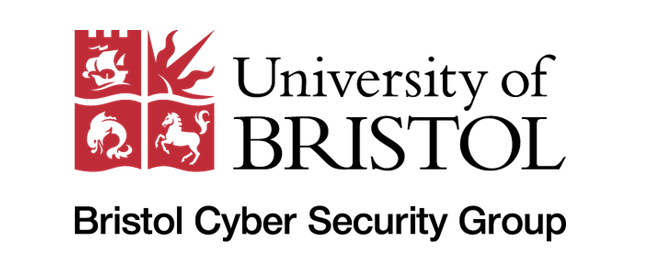}
\setlength{\fboxsep}{0pt}
\hfill \colorbox{cpiGray}{\makebox[3.22in][r]{\shortstack[r]{\vspace{2.75in}}}}%
\vspace{-0.25pt}
\setlength{\fboxsep}{10pt}
\setlength{\fboxrule}{0pt}
\colorbox{DarkRed}{\makebox[8.25in][l]{\hfill \shortstack[r]{\fontsize{30}{30}\rmfamily\color{white} A Survey of Relevant Text Mining Technology \\%
\fontsize{24}{24}\rmfamily\color{white}}}}%
\setlength{\fboxsep}{0pt}
\vspace{-8.5pt}
\hfill \colorbox{cpiGray}{\hspace{.25in} \parbox{2.97in}{\vspace{4in} \color{white} \textbf{Claudia Peersman \\ Matthew Edwards \\ Emma Williams \\ Awais Rashid \\\\   \today \vspace{2.3in} \vfill}}}%
\let\Title\title
\restoregeometry}

\mdfdefinestyle{SummaryBox}{%
    linecolor=DarkRed,
    outerlinewidth=3pt,
    roundcorner=20pt,
    innertopmargin=\baselineskip,
    innerbottommargin=\baselineskip,
    innerrightmargin=20pt,
    innerleftmargin=20pt,
    frametitlerule=true,
    frametitlerulecolor=DarkRed,
    frametitlefontcolor=white,
    frametitlebackgroundcolor=DarkRed,
    fontcolor=DarkRed,
    frametitle=Summary,
    backgroundcolor=gray!50!white,
    leftmargin=18pt,
    rightmargin=18pt}
    
\mdfdefinestyle{CharBox}{%
    linecolor=DarkRed,
    outerlinewidth=3pt,
    roundcorner=20pt,
    innertopmargin=\baselineskip,
    innerbottommargin=\baselineskip,
    innerrightmargin=20pt,
    innerleftmargin=20pt,
    frametitlerule=true,
    frametitlerulecolor=DarkRed,
    frametitlefontcolor=white,
    frametitlebackgroundcolor=DarkRed,
    fontcolor=DarkRed,
    frametitle=Characteristics,
    backgroundcolor=gray!50!white,
    leftmargin=18pt,
    rightmargin=18pt}

\mdfdefinestyle{MotivBox}{%
    linecolor=DarkRed,
    outerlinewidth=3pt,
    roundcorner=20pt,
    innertopmargin=\baselineskip,
    innerbottommargin=\baselineskip,
    innerrightmargin=20pt,
    innerleftmargin=20pt,
    frametitlerule=true,
    frametitlerulecolor=DarkRed,
    frametitlefontcolor=white,
    frametitlebackgroundcolor=DarkRed,
    fontcolor=DarkRed,
    frametitle=Motivations,
    backgroundcolor=gray!50!white,
    leftmargin=18pt,
    rightmargin=18pt}

\begin{document}

\begin{titlepage}
\maketitle
\end{titlepage}










\section{Introduction}
\label{sec:intro}

In recent years, Darknets and other environments offering anonymity are becoming increasingly popular among cybercriminals with a high degree of computer literacy and forensic awareness. Additionally, the emergence of online cybercriminal communities are enhancing the ``normalisation'' of cybercrime, providing offenders with technical and security support \cite{jeney2015combatting}. Although none of the existing anonymisation techniques (e.g., the ToR service (there are many legitimate and ethical uses of anonimisation techniques such as ToR, which we do not debate here (see \cite{troncoso2019privacy, dingledine2004tor, reed1998anonymous}) is entirely bulletproof, they can easily complicate or even block current cybercrime investigations. In such cases, the communications produced on social media platforms (both regular and cybercriminal fora) can be one of few clues to an offender's identity \cite{rocha2017authorship}. Additionally, investigating such social interactions can contribute to a better understanding of the dynamics leading to initial engagement in cybercrime, continued careers and (potentially) retirement.

Recent advances in text mining and natural language processing technology have enabled researchers to detect an author's identity or demographic characteristics, such as age and gender, in several text genres by automatically analysing the variation of linguistic characteristics. However, applying such techniques ``in the wild'' \cite{koppel2011authorship}, i.e., in both cybercriminal and regular online social media, differs from more general applications in that its defining characteristics are both domain and process dependent. This gives rise to a number of challenges of which contemporary research has only scratched the surface. More specifically, a text mining approach applied on social media communications typically has no control over the dataset size -- the number of available communications will vary across users. Hence, the system has to be robust towards limited data availability. Additionally, the quality of the data cannot be guaranteed. As a result, the approach needs to be tolerant to a certain degree of linguistic noise (for example, abbreviations, non-standard language use, spelling variations and errors). Finally, in the context of cybercriminal fora, it has to be robust towards deceptive or \textit{adversarial} behaviour, i.e. offenders who attempt to hide their criminal intentions (\textit{obfuscation}) or who assume a false digital persona (\textit{imitation}) \cite{brennan2009practical}, potentially using coded language.

In this work we present a comprehensive survey that discusses the problems that have already been addressed in current literature and review potential solutions. Additionally, we highlight which areas need to be given more attention. In the next section, we briefly introduce the fields of text mining and computational stylometry. Section \ref{sec:related} provides an overview of related work in these fields. In section \ref{sec:approach}, we discuss outstanding challenges and present the project's research agenda. Finally, we conclude this survey in Section \ref{sec:concl}. 

\newpage

\section{Background}
\label{sec:background}

\subsection{Text Mining}
\label{sec:BG_TM}
With the increasing availability of large amounts of computer-mediated communications, text mining has become a popular area of research for automatically detecting patterns and trends in a ``Big Data'' set-up. Typically designed within a Natural Language Processing (NLP) framework as a level of information extraction from text, the main objective of a text mining approach is to build an intelligent tool, that has the capability of analysing large amounts of natural language texts (e.g., newspaper articles, books or emails) and extracting useful information in a timely manner. Hence, it is a step forward from the information retrieval task, in which the best matches in a database are calculated based on a user query, to a level of exploring the various types of high quality knowledge that can be extracted from text. Although this is a relatively new research area, the technology is already being used in a wide variety of applications, such as biomedical applications (e.g., GoPubMed\footnote{\url{http://www.gopubmed.com/web/gopubmed/}}, a knowledge-based search engine for biomedical texts), business and marketing applications (e.g., stock return prediction \cite{galvez2017assessing}), security applications (e.g., automatic monitoring of Internet news, blogs and social media \cite{zanasi2009virtual}) and academic applications (e.g., academic publishers making their papers available for text mining purposes).

A text mining approach typically involves the following six steps:
\begin{enumerate}
\item A dataset of text documents relevant to the task at hand is collected. 
\item Each document is pre-processed. More specifically, the data is converted to the desired format, is split up into individual words and punctuation marks (i.e., \textit{tokenised}) and processed for removing content undesirable for the task in question (e.g., hyperlinks, non-standard word forms, redundancies and stop words).
\item The documents are transformed from the full text version to a vector space model that represents the different sets of linguistic features present in each document (e.g., words, characters, Parts-of-Speech and semantic roles). 
\item Statistical techniques are applied to determine which features are most informative for the task at hand. Usually, non-discriminative features are discarded by the system to reduce the dimensionality of the dataset.
\item The resulting structured database is analysed using either automatic classification or clustering techniques that are also used in data mining. In most cases, analysis happens using machine learning or statistical algorithms.
\item The output of the previous step is evaluated and can be stored or used in a series of following text mining experiments.
\end{enumerate}

\begin{figure}[h!]
\centering
\includegraphics{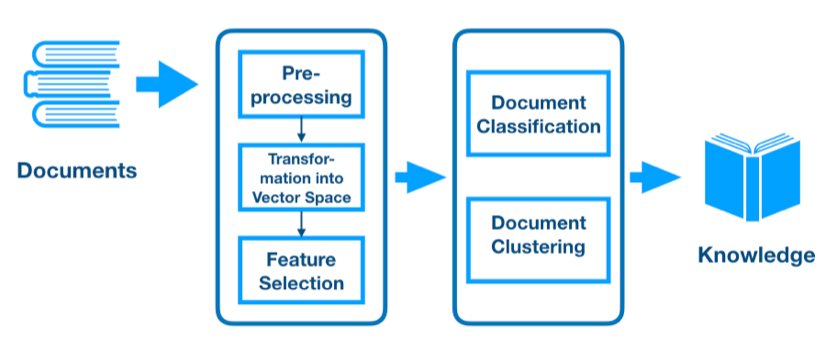}
\caption{A standard text mining process flow.}
\label{fig:TM_process}
\end{figure}

Figure \ref{fig:TM_process} shows a standard text mining process flow. The next section introduces the emerging research field of computational stylometry, in which the relation between natural language and its users is typically studied by adopting such a text mining framework.\\

\subsection{Computational Stylometry}
\label{sec:BG_CS}

Language is a social phenomenon and language variation is, as a consequence, innate to its social nature. By selecting between a range of potential variations of language use, people construct a social identity, which can be employed to achieve certain social goals. In other words, language users can make use of specific language varieties to represent themselves in a certain way. This freedom of choice, which is shaped by a series of both consciously and unconsciously made linguistic decisions, is often referred to as \textit{speaker agency}. Such variation can be manifested at various levels of language use, for example, the choice between different words, phonological variants or grammatical alternatives, and is typically influenced by a speaker's (intended) audience, demographic variables (such as age group, gender or background) and objectives (e.g., knowledge transfer, persuasion or likeability). Stylometry studies are mainly based on the hypothesis that the combination of these (un)conscious linguistic decisions is unique for each individual --- like a fingerprint or a DNA profile \cite{van2005new,daelemans2013explanation} --- and that language users (i.e., both speakers and authors) can be identified by analysing the specific properties of their linguistic choices. The idea of such a \textit{human stylome} \cite{van2005new} can be dated as far back as the mediaeval scholastics.

In modern times, the approach of analysing a text on different linguistic levels to determine its authorship was adopted within research fields such as stylistics and literary criticism, one of the most prominent examples being the investigations into the literary works attributed to Shakespeare \cite{dobson1992making}. This type of research is commonly referred to as \textit{traditional} authorship attribution and typically involves in-depth reading by human experts. However, in the late 19th century, a new line of research demonstrated that an author's writing style could be quantified. A study by \cite{mosteller1964inference}, for example, showed that the authorship of the Federalist Papers could be settled by comparing the distribution of \textit{stop words} (or \textit{function words}) in the disputed texts to other texts written by the three candidate authors\footnote{Alexander Hamilton, James Madison and John Jay.}. 

The arrival of modern computational methods and the emergence of the Internet have instigated a new research area, combining insights from the fields of stylometry to techniques commonly used in computer science. Based on the assumption that authors can be distinguished by their stylome, \textit{non-traditional} authorship attribution typically focuses on developing a computational model that can automatically identify the author of a given text. The dominant approach in these studies is typically based on text mining methods, which are used to automatically attribute one or more predefined thematic categories --- such as authors--- to a set of natural language texts (e.g., books, papers or emails)\footnote{The history and background of authorship attribution studies can be found in \cite{juola2008authorship}.}. Recently, a significant part of the field has shifted focus from attributing texts to specific authors to investigating whether certain aspects in an author's writing style can be generalised for larger author groups belonging to, for example, the same age or gender group or showing similar personality traits (e.g., outgoing or withdrawn). Together with non-traditional authorship attribution, such \textit{author profiling} studies constitute the rapidly developing field of computational stylometry. 

\section{Related Work}
\label{sec:related}

\subsection{Automatic User Profiling}
\label{sec:REL_profiling}
A large body of work already exists on detecting a user's age and gender based on computer-mediated communications. At first, most studies that involved a computational stylometry approach to automatically predict users' demographics were based on large collections of blogs (e.g., \cite{sarawgi2011gender,mukherjee2010improving,zhang2010predicting,goswami2009stylometric, argamon2009automatically, koppel2002automatically,nowson2007identifying,yan2006gender}. The main advantage of using blog corpora is that blog sites are publicly available and they usually contain information about the blogger's profile. In one such study, \citet{schler2006effects} applied a text mining approach to predict gender in a corpus of over 71,000 English blogs. Based on stylistic features (non-dictionary words, parts-of-speech, function words and hyper-links) and content features (content words\footnote{Content words carry the primary communicative message of an utterance (e.g., nouns, adjectives, verbs and adverbs).} with the highest Information Gain), they found that, despite the strong stereotypical differences in content between male and female bloggers, stylistic differences proved to be more discriminative than content differences \cite{schler2006effects}. However, combining both feature types, they were able to obtain an accuracy of 80.1\% when distinguishing between male and female bloggers. 

With regard to age (group) identification, content words showed to be slightly more useful than the style-based features, but again combining them rendered the best results \cite{schler2006effects}: 10s (13--17) were distinguishable from 30s (33--42) with accuracy above 96\% and differentiating between 10s and 20s (23--27) was achieved with an accuracy of 87.3\%. However, many 30s were wrongly classified as 20s, which rendered an overall accuracy of 76.2\%. This resulted in an F-score of 0.86 for the 10s, 0.75 for the 20s and 0.52 for the 30s category\footnote{These scores were calculated based on the confusion matrix in the paper.}. \citet{yan2006gender} were the first to include ``non-traditional'' features in their experiments, such as background colour, word fonts and cases, punctuation marks and emoticons. When combining these non-traditional features with bag-of-word features, their system achieved an F-score of 0.68 based on a corpus of 75,000 English blog entries authored by 3,000 individual bloggers. Interesting to see was that removing stop words actually decreased the performance of their system to 0.64, which is consistent with previous sociolinguistic studies that attested gender differences in the use of highly frequent word classes such as pronouns, articles and prepositions (e.g., \cite{mcmillan1977women,biber1998corpus,mulac2000female,mehl2003sounds,newman2008gender,keune2012explaining}). Similar results were found for age: based on the same corpus as was described in \cite{schler2006effects}, \citet{koppel2009computational} showed that language usage in blogs correlates with age: pronouns and the use of both assent and negation become scarcer with age, while prepositions and determiners become more frequent. Their system yielded an accuracy of 76.1\% for the three-way classification problem of attributing blogs to one of three age groups: 13--17, 23--27 or 33--47 (majority baseline = 42.7\%) by combining style- and content-based features and 80.5\% for predicting gender. \citet{goswami2009stylometric} further expanded the research of \cite{schler2006effects} by adding non-dictionary words and the average sentence length as features. Furthermore, the stylistic difference in usage of non-dictionary words combined with content words allowed to predict the age group (10s, 20s, 30s or higher) with an accuracy of 80.3\% and gender with an accuracy of 89.2\%. The average sentence length, however, did not correlate significantly with age or gender. Additionally, \cite{rao2010classifying} found that female authors were more likely to use emoticons, ellipses, character flooding, repeated exclamation marks, puzzled punctuation (i.e., combinations of ``?'' and ``!''), the abbreviation ``omg'' (\textit{oh my god}), and transcriptions of back-channels like ``ah'', ``hmm'', ``ugh'', and ``grr''. Affirmations like ``yeah'' and ``yea'' were the only preferences that were attributed to males. These latter features are called --- not quite accurately --- ``sociolinguistic features'' in e.g., \cite{rao2010classifying}. Finally, a number of other, non-textual features have been suggested for age and gender prediction, such as the number of friends and followers \cite{rao2010classifying, al2012homophily} and posted images \cite{yan2006gender}. 

More recently, a number of studies were based on a corpus of Twitter (e.g., \cite{nguyen2013old, bergsma2013using, fink2012inferring, bamman2012gender, al2012homophily, rao2010classifying}) and other social network data (see e.g., the author profiling tasks at PAN 2013, 2014 and 2015 \cite{rangel2013overview,rangel2014overview,rangel2015overview}). Although the amount of available data on Twitter is expanding massively, profile data is often absent, which requires additional techniques to acquire such meta-data. Contrary to blogs, tweets are typically very short, containing a maximum of 140 characters. However, most studies tend to combine multiple messages per user and show very similar results to previous studies on weblog data. The best results for gender prediction were achieved by \citet{bamman2012gender}, whose system achieved an accuracy score of 88.0\% based on over 600 tweets per user. When predicting age on a corpus of 200 Dutch tweets per user, \cite{nguyen2013old} were able to reach a 0.76 F-score when distinguishing between users younger than 20, between 20 and 40 years old and older than 40. Binary age prediction (adults versus adolescents), as examined in this chapter, was first performed by \citet{filippova2012user}, who investigated the performance of shallow textual features (e.g., character counts), language models and non-textual information (e.g., number of friends) when identifying bloggers under and over 18. However, their classifiers only yielded slightly better results than their majority baseline. Finally, \citet{rashid2013analyzing} presented a set of tools for predicting age and gender in a forensic context. By including POS, semantic and BOW features in a hierarchic classification system, their hierarchical, binary age prediction model yields probabilities that a user belongs to a specific age band (11--18 or over 18, followed by a breakdown of the probabilities for 11--14; 15--18; 19--49; 50+; etc.), resulting in a 72.15\% recall and 72.24\% precision for distinguishing between children and adults. 

Aside from investigating which feature types are most effective for predicting profile information, \citet{zhang2010predicting} contributed to the field by comparing different data representation methods, feature selection methods and machine learning algorithms for gender prediction in 3,226 blogs (52\% female), which contained about 400 words on average. They also included 20 semantic labels (e.g., ``conversation'', ``family'') as features in their instances, which were based on lists of words appearing in a similar context (e.g., ``tell'', ``talk'', ``ask'' belonged to the ``conversation'' label). Together with these word factor analysis features, they included word unigrams, POS tags and average word and sentence length in their experiments, but did not compare the results of these feature types individually. Their best prediction accuracy of 72.1\% was achieved by using Information Gain as feature selection criterion, and Support Vector Machines (linear kernel) as machine learning algorithm. Based on a corpus of 3,100 English blogs with an average post length of 250 words for men and 330 words for women, \cite{mukherjee2010improving} investigated which feature selection methods were most suitable for their type of data. Their ensemble feature selection method (EFS) improved the accuracy scores on gender attribution significantly compared to single selection metrics, such as Information Gain and Chi Square, by about 6-10\%, resulting in a best accuracy score of 88.6\%. Although this EFS method showed promising results, its application in age and/or gender attribution remains limited to \cite{mukherjee2010improving}. The reason for this could be that building a new classifier for each subset remains very time-consuming when working with a large number of features. 

Contrary to research on age and gender prediction, studies on automatically detecting a user's region of origin in social media communications are far less prominent in the field. \citet{rao2010classifying} experimented with token n-grams and the same set of sociolinguistic features that was mentioned above and were able to distinguish between English-writing Twitter users located in either Northern or Southern India with an accuracy score of 77.1\%. 

Although some of the previously mentioned studies show promising results for user profiling in social media communications, all of these works included text fragments ranging from 250 to several thousands of words on average per user. However, when looking at recent studies by \cite{peersman2018detecting,burger2011discriminating}, these results are subject to scalability issues when the models are applied on shorter text fragments: \citet{burger2011discriminating} reported a significant decrease in the performance to 66.5\% when predicting gender using only a single tweet per user. Additionally, the work of \cite{peersman2018detecting} specifically investigated the effect of different aspects of experimental design, such as feature types, feature selection, document representation and machine learning algorithms, when performing user profiling based on only one social media posting per user in the context of designing online child protection technology. The developed techniques will be evaluated for their efficiency when analysing cyber offenders' online messages in the project. 

\subsection{Adversarial Stylometry}
\label{sec:REL_adversarial}

Stylometry is based on the assumption that every individual has a unique writing style and, as a result, an author can be distinguished from other authors by measuring specific properties of his or her writings. However, most stylometric research is also based on the assumption that authors do not attempt to disguise their linguistic writing style. The author of \cite{juola2008authorship} discusses the importance of determining the robustness of an authorship attribution system when it is confronted with deception. However, so far, research into this issue has been limited. \citet{kacmarcik2006obfuscating} were the first to explore the possibility to computationally obfuscate the (most likely) author of the disputed Federalist Papers (see e.g., \cite{oakes2004ant}). They attempted to hide the author's identity by neutralising 14 of the most informative words per thousand words in the texts. Yet, the obfuscation was successfully detected by a technique called \textit{unmasking}, which was proposed by \cite{koppel2007measuring}: using a series of SVM classifiers to iteratively remove the features that received the highest weight from the SVM's during training, they found that, when comparing two texts that were written by two different authors, the accuracy score slowly declined during the iterations. However, when comparing two texts that were written by the same author of which one was computationally modified, as in \cite{kacmarcik2006obfuscating}, they attested a steep drop of the accuracy. This drop is explained by the fact that when comparing between texts that are written by the same author, the number of highly discriminative features is limited. Hence, when building a degradation curve, iteratively removing these features typically results in sudden drops in accuracy. However, when comparing texts that are written by different authors, the number of discriminative features is larger, resulting in a more steadily declining accuracy when iteratively removing subsets of these features (see also \cite{kestemont2012cross}). 

Contrary to these studies, however, initial work by \citet{brennan2009practical} showed that including obfuscation passages written by humans resulted in a devastating effect on the robustness of most state-of-the-art authorship attribution methods. Moreover, in an extended study on the English Brennan-Greenstadt corpus, which included original writings, obfuscated, and imitation passages of 45 different authors, the work of \cite{afroz2012detecting} stated that including obfuscation passages resulted in a decrease of the precision for authorship attribution from over 80\% to less than 10\% when training on data from forty different authors. With regard to detecting imitations of literary writings (or pastiches), \citet{dinu2012authorial} reported that using frequency rankings of stop words as features showed promising results when trying to distinguish between the Romanian novelist Caragiale's writings and authors that had attempted to imitate his writing style after his death. However, research by \cite{afroz2012detecting} reported a precision of less than 5\% when including imitation passages from the English Brennan-Greenstadt corpus. 

Although the work of \cite{juola2010empirical,juola2011analyzing} confirmed the fact that identifying the author of such deceptively written texts is extremely difficult, both \citet{afroz2012detecting} and \citet{juola2012detecting} found that the authors' intent to deceive or hide their identity is detectable. On the one hand, \citet{afroz2012detecting} reported that, in imitated passages, the usage of personal pronouns and particles increased, while the usage of adjectives decreased. They also noticed an increased use of existential ``there'', adverbs, particles and personal pronouns in obfuscated passages, but a decrease in the usage of nouns and wh-pronouns. Finally, they noticed that authors tend to ``dumb down'' their writing style by using shorter sentences, simpler words with less syllables, lower readability scores and higher readability ease and that changing function words seemed to be an important way to obfuscate a text. On the other hand, using the Java Graphical Authorship Attribution Program (JGAAP) software package, \cite{juola2012detecting} was able to identify five of six ``deceptive'' documents (83\%) and 22 out of 28 ``honest'' writings (79\%) in the Brennan-Greenstadt corpus. He concluded that the attempts of people to write ``differently'' could be fit into a recognisable and distinctive stylistic pattern.

Yet, the research described above still shows a number of limitations in the context of analysing cybercriminal social media communications: (a) some of the mentioned studies were performed on computationally modified data instead of on deceptive texts written by (untrained) human beings; (b) all studies were performed on a minimum of 500 words per author; and (c) they only included formal text genres. 

\newpage

\section{Towards Automatic User Profiling in Cybercriminal Communities}
\label{sec:approach}

\subsection{Challenges}

Three aspects are essential for designing a real-life computational stylometry application that can be used to support digital forensic investigations pertaining to cybercrime, namely: (\textit{i}) dealing with linguistically noisy texts, (\textit{ii}) sparse, skewed ``big data'' analysis and (\textit{iii}) detecting adversarial text samples. 

\textbf{Noisy data.} The increased level of immediacy in computer-mediated communication (CMC) has led to the rise of a new \textit{glocal} language variety, displaying characteristics from a global \textit{Internet language} leading to a wild proliferation of new language varieties (e.g., Internet abbreviations, acronyms, character flooding, concatenations and emoticons) \cite{androutsopoulos2004exploring, crystal2001language}. The presence of such linguistic noise is said to provide a significant challenge for text mining research, because many off-the-shelf NLP tools fail to correctly analyse this anomalous input. Previous work on 
age and gender detection in CMC discards all non-standard language varieties (e.g., \cite{schler2006effects,yan2006gender,nowson2007identifying}) or normalises them to improve feature extraction procedures (e.g., \cite{beaufort2010hybrid}). However, previous work in spoken discourse studies has observed strong correlations between the use of non-standard language and sociological variables such as age and gender (e.g., \cite{tagliamonte2012variationist,wolfram1969sociolinguistic,labov1972sociolinguistic,trudgill1983dialect,milroy1994glottal,cheshire2002sex,labov1990intersection,labov1994vol1,labov2001vol2,downes1998language,wardhaugh2010introduction,holmes2013introduction,chambers2013handbook}). Additionally, linguistic noise can also be used as an adversarial tactic by cybercriminals to avoid detection \cite{peersman2018detecting}. 

\textbf{Sparse, skewed Big Data.} Most documents only contain a small percentage of the total number of features present in the dataset. Because of their limited length, they provide great challenges for standard text mining approaches that rely on word frequencies, word co-occurrences or shared context to determine the similarity between documents. Additionally, in many cases the focus lies on detecting the minority class and, hence, the number of useful instances is limited. Standard practice in a wider text mining context is to increase the data in each sample by, e.g., grouping multiple text fragments written by the same author (\cite{schler2006effects,zhang2010predicting,nguyen2013old}) or by incorporating additional word level concept information obtained from external sources, such as pre-trained word embeddings, WordNet, concept annotations or snippets produced by public search engines (e.g., \cite{meng2013improving, heap2017word, liu2017leveraging}). 
However, for a real-life application, it is essential that a cyber offender profiling and detection approach is able to achieve a reliable performance, even when confronted with limited data availability. Furthermore, in a digital forensics context, it would be inconceivable to combine evidence with external content that was not produced by the person under investigation. 

\textbf{Adversaries.} Contemporary computational stylometry research typically focuses on two aspects: identifying and extracting linguistic features that are potentially discriminative for an author's \textit{writing print} (or \textit{stylome} \cite{van2005new}) and developing an efficient computational model that includes these features to automatically determine an author's identity or demographics. Although a range of feature types and computational methods have been suggested for the task, the field is dominated by studies that evaluate their computational stylometry approaches on non-deceptive datasets. However, a key issue when designing a computational stylometry approach to be used in cybercrime investigations is whether it will remain useful when it is confronted with adversarial behaviour. 
Cyber criminals may try to hide behind multiple digital personas or a group of offenders can share a single online identity. Additionally, they might attempt to hide their true identity or imitate other (non-criminal) users and use specialised vocabulary or coded language to conceal the nature of their activities \footnote{For example, illegal drug traffickers have been reported to use a widely varied terminology for selling their products \cite{nunn2010wanna}.}.

\subsection{Research Agenda}

In this section, we discuss the advances needed to study cybercriminal careers at scale. We focus on two key dimensions: (1) datasets that provide an empirically-grounded basis for uncovering identifying characteristics of cyber offenders; (2) technological advances for designing a text mining approach to automatically detect cybercriminal demographics and assessing such approaches for their robustness when applied in the wild and in adversarial settings of cybercriminal fora. 

\subsubsection{Datasets}

To support any text mining methodology, large and diverse datasets (Big Data) are required in order to study cybercriminal careers at scale. To our knowledge, the only significant longitudinal coverage of many cybercriminal communities is the dataset collected by \cite{dnmArchives}. This dataset consists of longitudinal observations of some 89 darknet marketplaces (crypto-markets using Bitcoin and other cryptocurrencies in escrow systems accessible as hidden services within the Tor network) and 37 forums over a period covering roughly 2013 to 2015, dependent on the marketplace accessibility. This large dataset, over 1.5TB of web pages and associated resources, contains within it a wealth of business and social interactions between criminals, covering several adverse events which had impact on the community, along with market listings, reputation indicators and other activity information and identity markers which are valuable for understanding the characteristics of offenders and their motivations. However, there are a number of limitations to this dataset, largely related to the incompleteness or partial inconsistency of particular crawling results. For example, pages may be missing from a given snapshot of a market on a particular date, due to scraper errors or connectivity issues with the market itself. These limitations can be partially addressed through the redundancy inherent in the longitudinal nature of the scraping process, inferring the approximate extent of missing data for a given site snapshot from prior and future observations of the same site, and translating this into imputed adjustments of the impact assessment measures. Other limitations include the comparatively smaller populations using darknet marketplaces for cyber-dependent crime as opposed to, e.g., illegal drug trade. There are, of course, other issues with the trustworthiness of observations from within the scrapes (e.g., vendors using shills, counterintelligence efforts, scams) and hold implications for any analysis. Finally, no (gold standard) information is available about the demographics of the users who are represented in the dataset.

For the purposes of the approach presented in the project, a new dataset with a broad coverage, which is being established by building a list of target criminal communities and collecting both archive and current data covering online communications between cyber criminals and adjacent participants, over as broad a period as possible in each case. Web-crawling technology is being deployed to unobtrusively collect online forum history, while conventional text-logging systems are used to monitor chat channels over a defined interval. This data will provide up-to-date information on the online language use of different types of cyber offenders, fuelling both the project's qualitative research objectives and text mining analyses and tool development. However, while this approach to corpus development would generate material suitable for unsupervised learning approaches (see below), it does not provide ``ground truth'' for the testing of user profiling models developed by the research team. 

Therefore, performance is being (partially) established through first establishing a baseline in similar non-criminal data for which ground-truth demographics are available:

\begin{itemize}
    \item The SMS-AP-18 corpus. This corpus was used for the first shared task on Multilingual Author Profiling on SMS (English/Urdu) \cite{fatima2018multilingual}. It consists of 810 user profiles together with their age metadata (15--19, 20--24, 25--xx) and gender.
    \item The PAN corpus (2012 -- 2017). This dataset contains different corpora collected from the author profiling tasks at PAN 2013, 2014 and 2015 \cite{rangel2013overview,rangel2014overview,rangel2015overview} and covers three online media genres (blogs, Twitter feeds and unspecified social media postings). All corpora used contain metadata about gender and age group (13--17, 23--27 and 33--47).
\end{itemize}

This will be followed by transfer learning and evaluation on smaller validated datasets established from criminal prosecutions or doxxing events, in which (anonymous) users are linked to their actual offline identity, drawn from criminal platforms themselves. We expect that such datasets will be too limited for training our systems, but will be valuable for evaluating the robustness of our approach when applied in the wild.

\subsubsection{Technological Advances}

\textbf{Computational Stylometry}. Our approach to this computational stylometry task is based on text categorisation, and involves the creation of document representations based on a selected set of (patterns) of linguistic features, feature selection using statistical techniques, and classification using machine learning algorithms (see Section \ref{sec:background}). Contrary to previous research that mainly focused on predicting authors' demographics based on large, formal text samples, we will perform a systematic study of different aspects of methodological design incorporated in a state-of-the-art user profiling approach to assess its robustness to highly sparse, skewed and noisy text data, i.e., performing user profiling ``in the wild''.

Based on the results of this systematic study, we will investigate different strategies to boost the performance for automatic user profiling using only a single message per user. Because, in the context of a cybercrime investigation, a conjunction of evidence with external content, which was not produced by the suspect\footnote{For example, by including information from Wikipedia, Wordnet or Internet search engines as is typically done in semantic and semi-supervised classification approaches.} would be out of the question, the following strategies will be explored:

\begin{enumerate}
\item \textbf{a systematic study} of different aspects of methodological design incorporated in a state-of-the-art user profiling approach to assess its robustness to highly sparse, skewed, noisy and adversarial text data, i.e., performing user profiling in the wild;
\item \textbf{novel feature engineering methods} in which different feature types are extracted in parallel and the resulting vectors are concatenated into larger vectors to create complex models; and
\item \textbf{a cross-task classification approach} in which the meta-data for gender or location is included in the experiments in order to investigate their effect on age prediction and vice versa.
\end{enumerate}

Additionally, we aim to investigate whether a \emph{message-based} approach, in which predictions are made on the level of the individual post and aggregated to the user level in post-processing, enables the system to identify different offender characteristics more accurately than the traditional \emph{user-based} approach that renders predictions directly on the user level. We will apply both approaches (a) in the context of automatically detecting people trying to imitate the writing style of another demographic group and (b) on the text genre of short, online social media communications. So far, both these applications have remained unexplored in the field.

Finally, the research project will also include a qualitative analysis of the features that make the user profiling model. Providing such a qualitative analysis of the most discriminative linguistic features enables an evaluation of the scalability of the approach for other researches and allows for methodological reflection towards well established sociolinguistic principles. Moreover, a qualitative analysis is typically absent in most text mining studies, as they tend to focus solely on the performance of their user profiling models.

\textbf{Topic Detection.} Aside from author profiling techniques, the project will focus on developing advanced unsupervised topic detection modules, which will be required for automatically analysing (underground) forums and marketplaces, allowing for automatically categorising and prioritising potential cyber threats and offences. Contrary to most text mining approaches, in which the training data available is pre-labelled with the required information to perform a categorisation task (i.e., the ``ground truth''), detecting topic in adversarial communications will require an \textit{unsupervised} learning approach, because no information on the presence or absence of guarded language will be available when analysing currently existing or newly collected datasets. As a result, methodologies will be investigated that can reveal hidden structures, patterns or features from unlabelled data. Potential machine learning techniques used in unsupervised learning that could contribute to this task include clustering techniques, Neural Networks and Formal Concept Analysis. 

\textit{Clustering}. In the baseline clustering approach, the similarity between different objects is measured by using one or more similarity functions. With regard to textual data in which the objects can be of different granularities (e.g., documents, paragraphs, sentences or words), clustering methods have shown promising results for e.g., browsing or organising documents and summarising large text corpora \cite{aggarwal2012survey}. Standard practice for vector data is to use the K-means algorithm or Latent Dirichlet Allocation (LDA). The first technique divides a set of text samples into \textit{k} disjoint clusters, each described by the centroid of the text samples in the cluster. The algorithm then attempts to select centroids that minimise the within-cluster sum-of-squares (or inertia) \cite{jain2010data}. LDA, from its part, is a Bayesian probabilistic model, which also assumes a collection of \textit{k} clusters. The latter algorithm could be especially useful when applied to social media communications, because it assumes that each document instance is a mixture of a small number of topics and that each word can be clustered into one of these topics \cite{blei2003latent}. Recent work on topic modelling over short text samples suggested incorporating word embeddings generated by Word2Vec \cite{mikolov2013efficient}. More specifically, short texts are aggregated into long pseudo-texts by incorporating the semantic knowledge from the word embeddings to boost the performance of clustering algorithms \cite{qiang2017topic}. A second challenge is that most clustering techniques depend on a predefined (\textit{k}) number of clusters. Therefore, agglomerative clustering algorithms will be investigated to determine the hierarchy of all topic clusters present in the dataset. More specifically, by using a bottom-up approach, in which each instance initiates its own cluster and clusters are merged together using a linkage criteria (for example, Ward’s algorithm \cite{ward1963hierarchical} hierarchical clustering can be achieved and represented in a tree structure or dendrogram for further analysis of the number of different topics present in the data.

\textit{Neural Networks}. Artificial neural networks can mainly be distinguished from other methods by its inclusion of one or more non-linear (or hidden) layers between the input and the output layer during the analysis. The input layer consists of a set of neurons, which represent the features of each instance. Next, each neuron in the hidden layer transforms the values from the previous layer with a weighted linear summation followed by a non-linear activation function. The output layer then analyses the values from the last hidden layer and produces a decision. In most cases, a supervised learning technique called Backpropagation is used during training, which runs a ``forward pass'' to compute all the activations throughout the neural network and determine the degree in which each node in each layer contributes to any errors in the output of the system \cite{pedregosa2011scikit, rumelhart1985learning}. In recent work by \cite{miao2017discovering}, neural topic models have been presented that show similar sparse topic distributions as found with traditional Dirichlet-Multinomial models on larger text samples, such as song lyrics or news articles. The advantage of recurrent neural networks, specifically, with regard to the task at hand is their ability to model sequences of unbounded length and, when combined with variational inference methods, they allow the number of topics to dynamically increase \cite{miao2017discovering}. 

\textit{Formal Concept Analysis}. FCA provides a well-founded mathematical framework for organising a set of objects based on their shared features without including any knowledge about the objects. In the context of automatic topic modelling in social media communications, the technique can be used to create thematically-based and cohesive clusters. The key advantage of FCA is that no prior knowledge of the data is required for its computation, which enables researchers to overcome typical topic detection problems, such as unknown topic distribution and the appearance of new topics. The technique has already been successfully applied on police data for detecting radicalisation and child sex offender grooming \cite{elzinga2011formalizing}.

\section{Conclusions}
Online interactions between cybercriminals are a valuable lens into the underlying nature of offenders, which in turn is necessary grounding for any preventative or disruptive intervention. Prior work has demonstrated some of the insight that studies of cybercriminal communities might have to offer~\cite{motoyama2011analysis,holt2013examining,decaryhetu2013reputation}. 
In this review, we described the related work, open challenges and requirements regarding computational assessment of cyber offenders, their identifying characteristics and their behaviours. These challenges and requirements underpin our research on analysing cybercriminal careers at scale. Within our on-going work in the project, 
we are focusing on developing a novel text mining approach for automatic user profiling and topic detection under the complex conditions of linguistically noisy, highly sparse, adversarial datasets and evaluate their forensic readiness when applied in the wild. Aside from addressing policy-guiding research questions, these new methodologies can be refined according to guidelines and feedback from law enforcement, leading to software tools that can support their investigations into cybercrime.

\label{sec:concl}





\bibliographystyle{model1-num-names}
\newpage\pagestyle{plain}
\bibliography{review.bib}


\end{document}